\documentclass[prl,twocolumn,floats]{revtex4}
\usepackage{mathrsfs}
\usepackage{graphicx}
\usepackage{amssymb}

\newcommand{\bea}[1]{\begin{eqnarray}\label{#1}}
\newcommand{\eea}{\end{eqnarray}}

\begin{document}
\title{Study of pseudogap and superconducting quasiparticle dynamics in $\rm{Bi_2Sr_2CaCu_2O_{8+\delta}}$ by time-resolved optical reflectivity}

\author{Xu-Chen Nie$^{1,2,3}$, Hai-Ying Song$^{3}$, Xiu Zhang$^{3}$, Yang Wang$^{3}$, Qiang Gao$^{4,5}$, Lin Zhao$^{6}$, Xingjiang Zhou$^{4,5,6}$, 
Jian-Qiao Meng$^{7}$, Yu-Xia Duan$^{7}$}
\author{Hai-Yun Liu$^{3,8}$}
  \email[Corresponding author:]{liuhy@baqis.ac.cn}
\author{Shi-Bing Liu$^{3}$}
  \email[Corresponding author:]{sbliu@bjut.edu.cn}

\affiliation{
\\$^{1}$Institute for Advanced Interdisciplinary Research, Nanjing University of Aeronautics and Astronautics, Nanjing 210016, China
\\$^{2}$College of Astronautics, Nanjing University of Aeronautics and Astronautics, Nanjing 210016, China
\\$^{3}$Strong-field and Ultrafast Photonics Lab, Institute of Laser Engineering, Beijing University of Technology, Beijing 100124, China
\\$^{4}$National Laboratory for Superconductivity, Beijing National Laboratory for Condensed Matter Physics, Institute of Physics, 
            Chinese Academy of Sciences, Beijing 100190, China
\\$^{5}$University of Chinese Academy of Sciences, Beijing 100049, China
\\$^{6}$Collaborative Innovation Center of Quantum Matter, Beijing 100871, China
\\$^{7}$School of Physics and Electronics, Central South University, Changsha, Hunan 410083, China
\\$^{8}$Beijing Academy of Quantum Information Sciences, Beijing 100193, China
}

\begin{abstract}
{\bf  The relation between pseudogap (PG) and superconducting (SC) gap, whether PG is a precursor of SC or they coexist or compete, is a long-standing controversy in cuprate high-temperature supercondutors. Here, we report ultrafast time-resolved optical reflectivity investigation of the dynamic densities and relaxations of PG and SC quasiparticles (QPs) in the underdoped $\rm{Bi_2Sr_2CaCu_2O_{8+\delta}}$ ($T_c$ = 82 K) single crystals. We find evidence of two distinct PG components in the positive reflectivity changes in the PG state, characterized by relaxation timescales of $\tau_{fast}$ $\approx$ 0.2 ps and $\tau_{slow}$ $\approx$ 2 ps with abrupt changes in both amplitudes $A_{fast}$ and $A_{slow}$ at the PG-opening temperature $T^*$. The former presents no obvious change at $T_c$ and coexists with the SC QP. The latter's amplitude starts decreasing at the SC phase fluctuation $T_p$ and vanishes at $T_c$ followed by a negative amplitude signifying the emergence of the SC QP, therefore suggesting a competition with superconductivity.}
\end{abstract}
\date{\today}
\maketitle
\section{Introduction}
It is well-known that hole-doped cuprate high-$T_c$ superconductors (HTSCs) exhibit an unusual normal state that is characterized by the opening of the pseudogap (PG) in the electronic density of states (DOS), at a temperature $T^*$ well above the superconducting (SC) transition temperature $T_c$ \cite{Ding1996Spectroscopic,Loeser1996Excitation,Marshall1996Unconventional,Keimer2015From}. The PG affects electronic structures by obliterating the antinodal regions of the Fermi surface near the Brillouin zone edge, leaving only ``Fermi arcs" around the nodes \cite{Renner1998Pseudogap,Norman1998Destruction}. The properties of the PG and its relation with SC gap are important in cuprate high-temperature superconductors. One of the central puzzles in cuprate HTSCs lies on whether PG acts as a precursor to the SC gap pairing (known as ``one-gap" picture) \cite{Kanigel2007Protected,Shi2008Coherent,Meng2009Monotonic} or originates from other independent/competing orders (known as ``two-gap" picture) \cite{Tanaka2006Distinct,Kondo2007Evidence,Lee2007Abrupt,Terashima2007Anomalous,Kondo2009Competition}. Recent scanning tunneling microscopy (STM) and resonant elastic X-ray scattering (REXS) experiments have suggested that charge ordering is responsible for the PG opening and competes with superconductivity \cite{Comin2014Charge,da2014Ubiquitous}. Competition between PG and SC order parameters is further confirmed by angle-resolved photoemission spectroscopy (ARPES), in which a singularity in the spectral weight has been observed at $T_c$ at the antinode \cite{Hashimoto2014Energy,Hashimoto2015Direct}.

In time-resolved optical reflectivity measurements, a first femtosecond (fs) laser pulse pumps the system into a non-equilibrium state, subsequently a relatively weaker laser pulse is implemented to probe the reflectivity change $\Delta R/R$, which is used to infer the intrinsic charge dynamics, based on the relation $\Delta R = (\partial R/\partial n)n_{pe}$ when the photoexcited quasiparticle (QP) density ($n_{pe}$) is small compared to the thermally excited one\cite{Demsar2001Quasiparticle}. By varying the pump-probe delay, the relaxation process in time-domain is obtained. In HTSC and charge density wave (CDW) systems, the relaxation process is bottlenecked by the presence of the charge gaps when the photoexcited QPs relax to states near the Fermi level. Subsequently, the photoexcited QPs recombine by the emission and reabsorption of bosons with energy greater than 2$\Delta$
\cite{Demsar2001Quasiparticle,Demsar1999Superconducting,Kabanov1999Quasiparticle,Nair2010Quasiparticle,Coslovich2011Evidence,Toda2011Quasiparticle,Giannetti2016Ultrafast,Demsar1999Single,Shimatake2007Selective,Stojchevska2010Electron,Pogrebna2015Superconducting,Tian2016Ultrafast,Chu2017A,Boschini2018Collapse,Konstantinova2018Nonequilibrium}. Therefore this method allows to directly distinguish the gap character by tracking relaxation processes. One outstanding example is that the PG and SC QPs in cuprate superconductors have been found to separately relax on different timescales and coexist below $T_c$, forming two-component exponential relaxations in $\Delta R/R$ curves below $T_c$ and single-component exponential relaxation above $T_c$ 
\cite{Nair2010Quasiparticle,Toda2011Quasiparticle,Coslovich2013Competition,Liu2008Direct,Toda2014Rotational,Vishik2017Ultrafast}. However, comprehensive ultrafast study of the competition between PG and SC QP dynamics is still lacking.

In this paper, we use ultrafast time-resolved optical reflectivity to investigate the QP dynamics in both the PG and SC states of the underdoped $\rm{Bi_2Sr_2CaCu_2O_{8+\delta}}$ (UD-Bi2212, $T_c$ = 82 K, $T^*$ $\approx$ 180 K) single crystals. The PG state consists of two PG components: a fast one with $\tau_{fast}$ $\approx$ 0.2 ps and a slow one with $\tau_{slow}$ $\approx$ 2 ps, corresponding to two types of PG QPs, giving rise to two-component exponential relaxations in the transient reflectivity changes $\Delta R/R$ above $T_c$. Both the fast and slow PG QPs show abrupt changes around $T^*$. The fast PG QP presents no obvious changes in its amplitude $A_{fast}$ and relaxation $\tau_{fast}$ at $T_c$, and coexists with the SC QP. On the other hand, the amplitude $A_{slow}$ of the slow PG QP significantly drops below the SC phase fluctuation temperature $T_p$, and then vanishes at $T_c$ followed by a reversed sign in $\Delta R/R$ signifying the emergence of the SC QP. A sharp divergence in $\tau_{slow}$ is obviously evidenced at $T_c$, confirming the competition between the slow PG and the SC dynamics. Further analysis by using the Rothwarf-Taylor model indicate that the fast PG QP is well fitted by a $T$-independent gap, while the slow PG QP is better fitted by a $T$-dependent BCS-like gap, suggesting a charge order nature that competes with superconductivity. Our fittings yield a SC gap $\Delta_{SC}(0)$ = 17 $\pm$ 1 meV and two PG gaps $\Delta_{PGfast}$ = 48 $\pm$ 2 meV and $\Delta_{PGslow}(0)$ = 116 $\pm$ 17 meV, consistent with the SC gap nearby the node, the antinodal PG and the high-energy hump feature at the antinode, respectively. Our studies indicate that time-resolved technique is an effective way of tracking the PG and SC QPs, and unveil their relations in cuprate HTSCs.

\section{Results}
Our ultrafast time-resolved optical reflectivity measurements were carried out on single crystals of UD-Bi2212 using optical
pulses with temporal duration of 35 fs and centre wavelength of 800 nm, produced by a regeneratively amplified Ti:sapphire laser system
operating at a repetition rate of 1 kHz, as described in \cite{Nie2018Transient,Nie2018Photoinduced}. The pump and probe pulses were focused onto the sample at near-normal incidence with spot diameters of $\sim$ 0.4 mm and $\sim$ 0.2 mm, respectively, which ensured an excellent pump-probe
overlap. With the pump beam modulated by a mechanical chopper, the reflected probe signal was
focused onto a Si-based detector which was connected to a lock-in amplifier, where the photoinduced changes in reflectivity
$\Delta R/R$ were recorded. The time evolution
of the pump-induced changes in the probe reflectivity ($\Delta R/R$) were measured by scanning the delay time between pump and
probe pulses, using a motorized delayline. To perform the temperature-dependent measurements, the sample was mounted on a
cryostat with a temperature sensor embedded close by, allowing a precise control of temperature in the range of 5$-$325 K. The pump fluence ($F$) was tunable between 12 and 300 $\mu\rm{J/cm^2}$ by using neutral density filters. The probe fluence of 4 $\mu\rm{J/cm^2}$ was chosen to minimize any probe-induced perturbation of the system.

\begin{figure}[tbp]
\begin{center}
\includegraphics[width=1\columnwidth,angle=0]{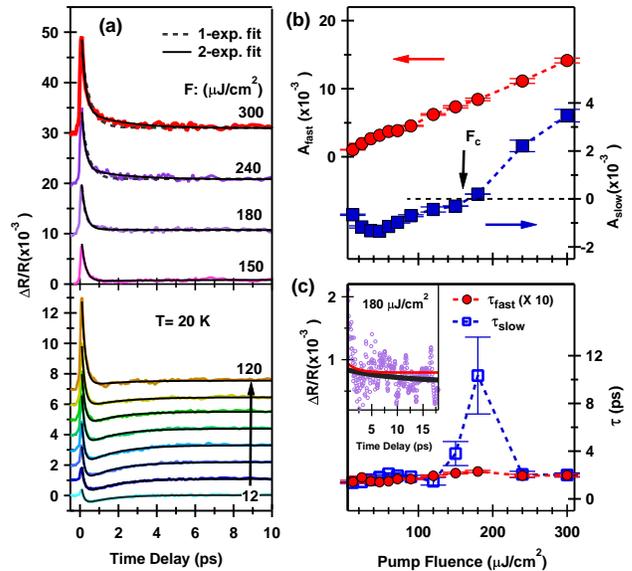}
\end{center}
\caption{{\bf Photoinduced melting of the SC phase.} ({\bf a}) Time-resolved reflectivity changes $\Delta R/R$ measured at 20 K, with pump fluence in the range from 12 to 300 $\mu\rm{J/cm^2}$. The dashed and solid black curves are single-component and two-component exponential fits, respectively. Fluence-dependent amplitudes ({\bf b}) and relaxations ({\bf c}) of QP dynamics are extracted from ({\bf a}) using two-component exponential fits. The error bars are obtained from fittings. The black arrow marks the critical fluence $F_c$ where the photoinduced melting of the SC phase occurs and the slow decay emerges. Inset: Zoom-in view of the slow relaxation in 180 $\mu\rm{J/cm^2}$ with nondivergent $\tau_{slow}$ (2 ps, red solid curve) and divergent $\tau_{slow}$ (10.3 ps, black solid curve), respectively.}
\end{figure}
Figure 1 shows reflectivity changes as a function of pump fluence $F$ at 20 K. In our cases the pump and probe polarizations were both along the $a$-axis of the sample crystal, in order to separate the PG and SC dynamics with different signs and timescales as observed in \cite{Liu2008Direct}. The $\Delta R/R$ curves below $F_c$ ($\approx$ 160 $\mu\rm{J/cm^2}$) obviously consist of a fast relaxation with positive amplitude and a slow relaxation with negative amplitude, corresponding to the fast PG QP and SC QP respectively, in good agreement with previous studies \cite{Liu2008Direct,Toda2014Rotational}. Above $F_c$ the slow component presents a positive sign following the disappearance of the negative relaxation. Therefore, it is reasonable to use a two-component exponential function to describe all the relaxation processes: $\Delta R/R(t)=A_{fast}exp({-t/{\tau_{fast}}})+A_{slow}exp({-t/{\tau_{slow}}})+A_0$, where $A_{fast}(\tau_{fast})$ and $A_{slow}(\tau_{slow})$ represent the amplitudes (relaxations) of the fast and slow components respectively, and $A_0$ describes much far slower equilibration processes out of the period of the measurement, such as heat diffusion. As shown in figures 1(b) and (c), $A_{fast}$ is linearly proportional to $F$, and $\tau_{fast}$ is constant at $\sim$ 0.2 ps, suggesting that the fast PG is $F$-independent and the bottleneck-mechanism is available for the relaxation in our measurement regime. $A_{slow}$ is more closely $F$-dependent and consists of three regimes: (1) below 50 $\mu\rm{J/cm^2}$, $A_{slow}$ linearly increases with $F$; 
(2) $A_{slow}$ saturates and decreases with $F$ increasing from 50 $\mu\rm{J/cm^2}$ to $F_c$; and (3) $A_{slow}$ reverses to positive and linearly increases with $F$ above $F_c$.

\begin{figure}[tbp]
\begin{center}
\includegraphics[width=1\columnwidth,angle=0]{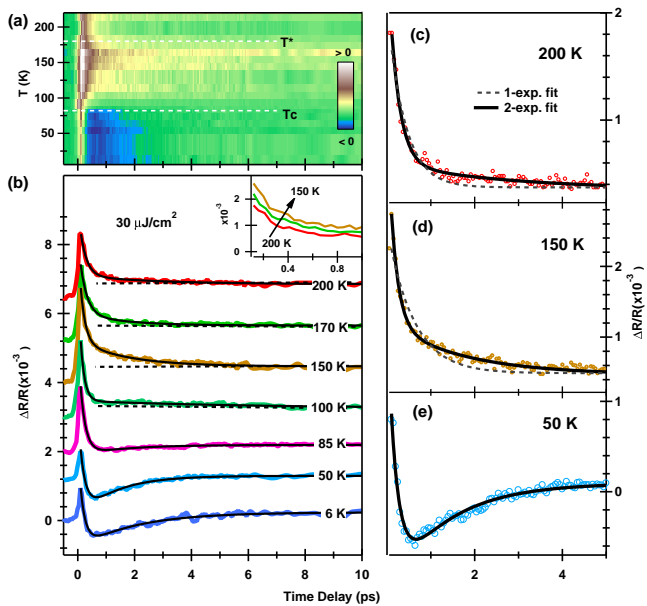}
\end{center}
\caption{{\bf Temperature-dependent reflectivity changes of UD-Bi2212 ($T_c$ = 82 K).} ({\bf a}) False-color image showing $\Delta R/R$ as a function
of time delay and temperature for an underdoped Bi2212 sample with $T_c$ = 82 K and $T^*$ $\approx$ 180 K (white dashed lines). ({\bf b}) Time-resolved reflectivity changes $\Delta R/R$ measured at 30 $\mu\rm{J/cm^2}$ over a range of temperatures. The inset depicts the evolution of the fast decay around $T^*$. The horizontal dashed lines are used to highlight the slow decay feature above $T_c$. ({\bf c})-({\bf e}) Representative curves of the reflectivity changes (open circles) at temperatures 200, 150 and 50 K, with single-component (dashed lines) and two-component exponential fits (solid lines).}
\end{figure}

To study the intrinsic dynamics of the PG and SC QPs and their relations, and to make sure the bottleneck-mechanism (constant relaxations) and the relation $\Delta R=(\partial R/\partial n) n_{pe}$ available (weak photoexcitation), we carried out temperature-dependent measurements at the pump fluence 30 $\mu\rm{J/cm^2}$, as shown in figure 2.
The color changes in figure 2(a) correspond to the variation of QP densities across $T_c$ and $T^*$, marked by white dashed lines. Obviously, a slow relaxation with positive amplitude appears below $T^*$, and changes to negative below $T_c$. Time-dependent curves in figure 2(b) provides more details that the data below $T_c$ obviously consist of a positive $A_{fast}$ and a negative $A_{slow}$, and above $T_c$, $A_{slow}$ switches to positive, analogous to the curves above $F_c$ in figure 1(a). This scenarios again confirm that $A_{slow}$ with negative sign represents the SC QP dynamics, and suggest that $A_{slow}$ with positive sign above $T_c$ refers to the PG QP dynamics. In figures 2(c)-(e), we chose three typical $\Delta R/R$ curves at 200, 150 and 50 K with two-component exponential fits and as a comparison, we also plot single-component exponential fits. Obviously, only two-component exponential fits can reproduce all measured curves, while single-component exponential fits are not able to match especially the slow relaxation.

\begin{figure*}[tbp]
\begin{center}
\includegraphics[width=2\columnwidth,angle=0]{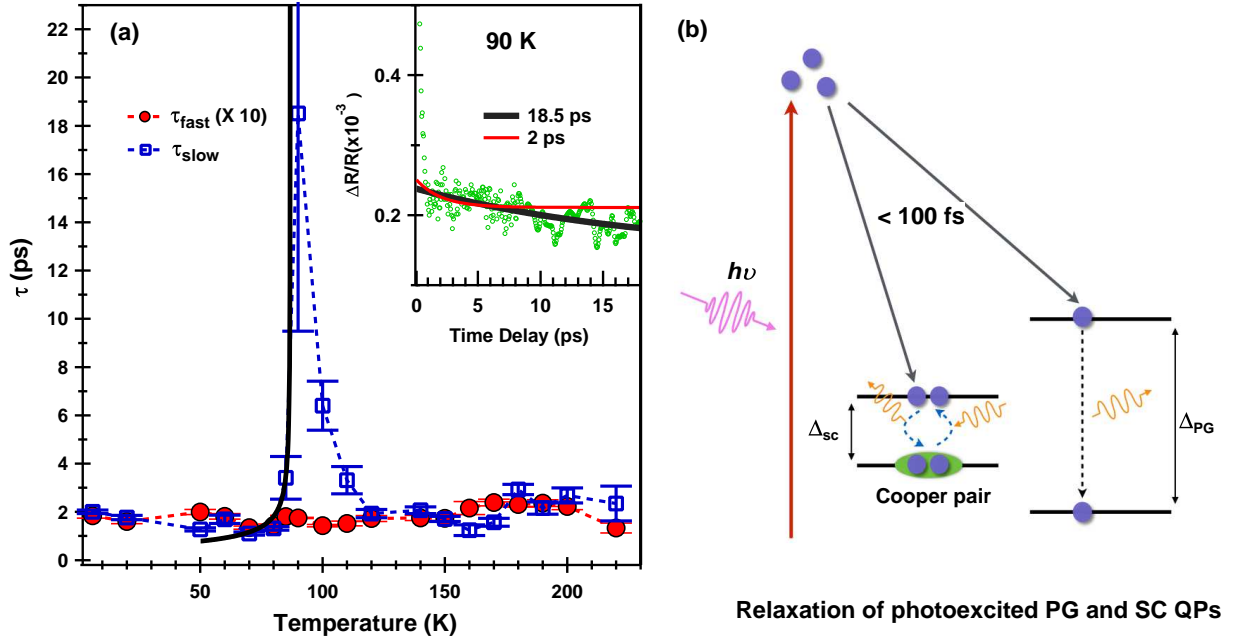}
\end{center}
\caption{{\bf  Temperature evolution of the relaxations.} ({\bf a}) The relaxations of the QP dynamics (${\tau_{fast}}$ and ${\tau_{slow}}$)
as a function of temperature are extracted from figure 2 by two-component exponential fits. The error bars are obtained from the fits. The black curve is a fit nearby $T_c$ using $\tau \propto 1/\Delta(T)$ with a BCS-like $T$-dependent gap. Inset: Zoom-in view of the slow relaxation at 90 K with nondivergent $\tau_{slow}$ (2 ps, red solid curve) and divergent $\tau_{slow}$ (18.5 ps, black solid curve), respectively.
 ({\bf b}) A schematic illustration of the relaxation bottleneck of photoexcited SC and PG QPs. After the initial fast electron-electron thermalization and avalanche energy relaxation, the photoexcited QPs temporally redistribute on the gap edges. The SC QPs can recombine into Cooper pairs with the emission of high-frequency bosons (with energy $\hbar\omega>2\Delta_{SC}$), which can immediately break Cooper pairs, leading to a cyclical process as illustrated by the blue dashed arrows. On the other hand, the PG QPs recombine through the emission of bosons.}
\end{figure*}

In figure 3(a) we plot the relaxations $\tau_{fast}$ and $\tau_{slow}$ as a function of temperature. $\tau_{fast}$ is found to be $T$-independent without any obvious change around $T_c$, and in contrast, $\tau_{slow}$ is $T$-dependent with a sharp divergence nearby $T_c$. Figure 3(b) illustrates the relaxation process of the photoexcited QPs at SC phase. After the initial photoexcitation, fast electron-electron thermalization and avalanche energy relaxation last only few tens of femtoseconds \cite{Giannetti2016Ultrafast,Boschini2018Collapse,Konstantinova2018Nonequilibrium}, giving rise to a unresolvable sharp rising edge in $\Delta R/R$ because it is comparable to the laser pulse duration. As such photoexcited QPs from high excited state far above the Fermi level relax to the upside of the PG and SC gap edges with a relaxation bottleneck. Then the photoexcited PG and SC QPs recombine through interaction with high-frequency bosons in the equilibration process, corresponding to the two-component exponential relaxations in our case.

\section{Discussion}
For the superconducting phase transition or, more generally, any kind of symmetry-breaking transition, the opening of a gap in the electronic density of states (DOS) close to the Fermi level introduces additional phase-space constraints for the scattering processes, which typically results in a bottleneck effect in the relaxation dynamics. For an electron-boson coupling system, the recombination of the photoexcited SC QP as a function of temperature, can be described by the Rothwarf-Taylor model phenomenologically \cite{Rothwarf1967Measurement}, in which bosons with energy higher than 2$\Delta$ dominate the recombination process. When $T$ approaches $T_c$, the gap closes and more low-energy bosons are involved for the relaxation processes, resulting in a divergence nearby $T_c$, which can be simply described to be proportional the inverse of the gap as $\tau(T) \propto 1/{\Delta(T)}$ \cite{Kabanov1999Quasiparticle}. Figure 3(a) shows a fit to $\tau_{slow}$ in the vicinity of $T_c$ using a BCS-like $T$-dependent gap $\Delta(T)=\Delta(0)\sqrt{1-T/T_c}$, where $\Delta(0)$ is the gap at 0 K. It has already been found that the divergence of SC/CDW QPs relaxation nearby $T_c$ is a universal feature in cuprates
\cite{Kabanov1999Quasiparticle,Demsar1999Single,Demsar2001Quasiparticle,Nair2010Quasiparticle,Coslovich2011Evidence,Toda2011Quasiparticle,Giannetti2016Ultrafast,Liu2008Direct}, pnictides \cite{Stojchevska2010Electron,Pogrebna2015Superconducting,Tian2016Ultrafast} and CDW systems \cite{Demsar1999Single,Shimatake2007Selective,Chu2017A}.

\begin{figure}[tbp]
\begin{center}
\includegraphics[width=1\columnwidth,angle=0]{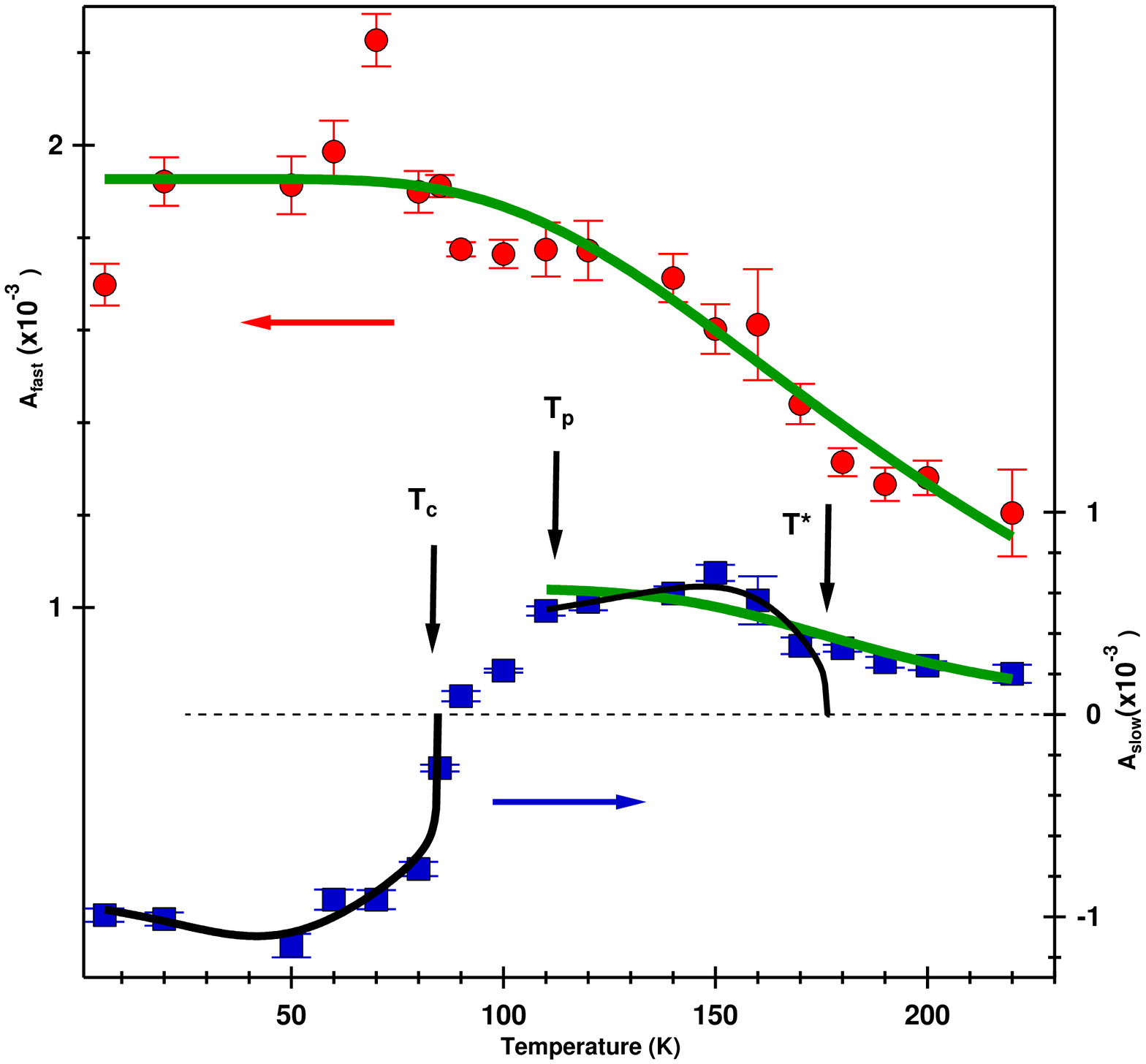}
\end{center}
\caption{{\bf Temperature evolution of the amplitudes.} The amplitudes ${A_{fast}}$
(red circles) and ${A_{slow}}$ (blue squares) as a function of temperature, extracted from figure 2 by two-component exponential fits. The error bars are obtained from the fits. The black curve is fitted by (1) with a BCS-like $T$-dependent gap, while the green curves are fits by (2) with a $T$-independent gap.}
\end{figure}

Using the same model, the $T$ dependence of the SC component $A_{slow}$ in figure 4, can be simulated as \cite{Kabanov1999Quasiparticle}
\bea{}
|A(T)|\propto n_{pe}=\frac{\epsilon_I/(\Delta(T)+k_BT/2)}{1+B\sqrt{\frac{2k_BT}{\pi\Delta(T)}}exp(-\Delta(T)/k_BT)}
\eea
where $\epsilon_I$ is the pump laser intensity per unit cell, $B = 2\nu/N(0)\hbar\Omega_c$ is the parameter determined by $\nu$ the number of boson modes per unit cell, in which $N(0)$ is the density of states at the Fermi level and $\Omega_c$ is the boson cutoff frequency. $\Delta(T)$ is a BCS-like gap which is proven to be applicable in time-resolved optical reflectivity from underdoped to overdoped cuprates \cite{Demsar1999Superconducting,Kabanov1999Quasiparticle,Nair2010Quasiparticle,Coslovich2011Evidence,Toda2011Quasiparticle,Giannetti2016Ultrafast}. From the analysis above, we obtain $\Delta_{SC}(0)$ = 17 $\pm$ 1 meV, in good agreement with the values by two-color pump-probe spectroscopy \cite{Toda2011Quasiparticle}, scanning tunneling microscopy/spectroscopy (STM/STS) \cite{Kurosawa2010Large} and SC gap in the nodal region by ARPES\cite{Tanaka2006Distinct,Lee2007Abrupt,Hashimoto2014Energy,Hashimoto2015Direct}.

For the case of the fast PG, which is $T$-independent, then the relaxation amplitude $A_{fast}$ is given by \cite{Kabanov1999Quasiparticle}
\bea{}
|A(T)|\propto n_{pe}=\frac{\epsilon_I/\Delta_{PG}}{1+Bexp(-\Delta_{PG}/k_BT)}
\eea
The fitting by (2), as shown in figure 4, presents a good agreement with the fast PG component, yielding $\Delta_{PGfast}$ = 48 $\pm$ 2 meV, corresponding to the antinodal pseudogap as observed by ARPES \cite{Kondo2007Evidence,Lee2007Abrupt}.

The slow PG component $A_{slow}$ around $T^*$ obviously can not be reproduced by (2) with a $T$-independent gap, as depicted by the green curves in figure 4. Rather than that $A_{slow}$ below $T^*$ can be well fitted by (1) with a BCS-like gap, yielding a fitted value $\Delta_{PGslow}(0)$ = 116 $\pm$ 17 meV. $A_{slow}$ between $T_c$ and $T^*$ is closely dependent on temperature: it first arises at $T^*$, suggesting its PG nature; then it starts to decrease sharply at $T_p$ = 110 K, consistent with the SC phase fluctuation temperature well observed by time-resolved ARPES \cite{Zhang2013Signatures}; finally, it is completely suppressed to zero nearby $T_c$ and upon further cooling turns to a reversed sign which is an indication of the SC QP, accompanied by a divergence in $\tau_{slow}$ as observed in figure 3(a). 

As for the origin of the slow relaxation between $T_c$ and $T^*$, we can exclude any in-gap or impurity-induced intermediate states that affect and extend the relaxation process, because of its $F$- and $T$-dependent behaviors. The fitted value $\Delta_{PGslow}(0)$ = 116 $\pm$ 17 meV, is consistent with the high-energy hump feature at the antinode that is strongly affected by the PG \cite{Hashimoto2015Direct}. Then the decrease of $A_{slow}$ with temperature decrease around $T_c$ might be linked to the suppression of the antinodal spectral weight of the PG order which competes with the SC order \cite{Hashimoto2015Direct}.

\section{Conclusion}
Both $A_{fast}$ and $A_{slow}$ present an increase with temperature decrease around $T^*$ (figure 4), indicating the emergence of two types of PG QPs: a fast and a slow one, which can be described by the Rothwarf-Taylor model with a $T$-independent gap and a BCS-like gap, respectively. It is also worth to mention that both $A_{fast}$ and $A_{slow}$ remain observable above $T^*$, which maybe related to electron-boson coupling and PG fluctuation \cite{Chia2013Doping}, in contrast to a sharp drop to zero in $A_{slow}$ nearby $T_c$. These scenarios suggest the possible existence of PG fluctuations above $T^*$, analogous to the fluctuation of the CDW order above the transition temperature in blue bronze \cite{Demsar1999Single}.

In summary, we have investigated the fluence- and temperature-dependent dynamics of the PG and SC QPs in UD-Bi2212 by ultrafast time-resolved optical reflectivity. Our results clearly indicate that two types of PG QPs, the fast and slow PG QPs, coexist in the PG state. The fast PG QP coexists with the SC QP on distinct timescales below $T_c$. The slow PG QP is completely suppressed at $T_c$ and replaced by the SC QP with reversed sign in the slow relaxation component, suggesting a competition between them.

{\bf Acknowledgments} We gratefully acknowledge financial support from the National Natural Science Foundation of China (Grant No. 51705009) and the NSAF of China (Grant No. U1530153). HYL thanks the Key Project of Beijing Municipal Natural Science Foundation and Beijing Education Committee Science and Technology Plan (Grant No. KZ201810005001) and the Sea Poly Project of Beijing Overseas Talents (SPPBOT). XJZ thanks financial support from the National Key Research and Development Program of China (2016YFA0300300), the National Natural Science Foundation of China (Grant No.11334010), and the Strategic Priority Research Program (B) of the Chinese Academy of Sciences (Grant No. XDB07020300).

{\bf Conflict of interest} The authors declare that they have no conflict of interest.

\bibliographystyle{aipnum4-1}
\bibliography{Bi2212}

\end{document}